\documentclass[12pt,letter]{article}
\usepackage[latin9]{inputenc}
\usepackage{color}
\usepackage{verbatim}
\usepackage{float}
\usepackage{amsthm}
\usepackage{amsmath}
\usepackage{amssymb}
\usepackage{graphicx}
\usepackage{setspace}
\usepackage[authoryear]{natbib}
\usepackage[unicode=true,
 bookmarks=true,bookmarksnumbered=false,bookmarksopen=false,
 breaklinks=false,pdfborder={0 0 1},backref=false,colorlinks=true]
 {hyperref}
\hypersetup{
 pdfauthor={Xi Luo},
 citecolor=blue}
\usepackage{breakurl}

\makeatletter

\floatstyle{ruled}
\newfloat{algorithm}{tbp}{loa}
\providecommand{\algorithmname}{Algorithm}
\floatname{algorithm}{\protect\algorithmname}

\theoremstyle{plain}
\newtheorem{thm}{\protect\theoremname}
  \theoremstyle{plain}
  \newtheorem{prop}[thm]{\protect\propositionname}

\usepackage{amsfonts}\usepackage{amsbsy}\usepackage{epsfig}\@ifundefined{definecolor}
 {\usepackage{color}}{}
\usepackage{rotating}

\makeatother

  \providecommand{\propositionname}{Proposition}
\providecommand{\theoremname}{Theorem}

\begin{document}
\noindent \begin{center}
\textbf{\large{A Hierarchical Graphical Model for Big Inverse Covariance
Estimation with an Application to fMRI}} \textbf{\large{\vspace*{-40pt}
}}
\par\end{center}{\large \par}

\begin{singlespace}
\begin{flushleft}
\noindent \begin{center}
Xi Luo\\
Brown University\textbf{\large{\vspace*{-30pt}
}}
\par\end{center}
\par\end{flushleft}
\end{singlespace}

\begin{center}
\let\thefootnote\relax\footnote{Xi Luo is Assistant Professor, with Department of Biostatistics, Center for Statistical Sciences, and Computation in Brain and Mind Initiative, Brown University, Providence, RI 02912. Date: \today. \\ \hspace*{1.8em} XL would like to acknowledge partial support from National Institutes of Health grants  P01AA019072, P20GM103645,  P30AI042853,  R01NS052470, and S10OD016366, a Brown University Research Seed award, a Brown Institute for Brain Science Pilot award, and a Brown University faculty start-up fund.\\ \hspace*{1.8em} This paper has been presented orally at Yale University on Feburary 18, 2014, and at the Eastern North American Region Meeting of the International Biometric Society on March 18, 2014.\\ \hspace*{1.8em} An R package of the proposed method will be publicly available on CRAN.} 
\par\end{center}

\textbf{Abstract}. Brain networks has attracted the interests of many
neuroscientists. From functional MRI (fMRI) data, statistical tools
have been developed to recover brain networks. However, the dimensionality
of whole-brain fMRI, usually in hundreds of thousands, challenges
the applicability of these methods. We develop a hierarchical graphical
model (HGM) to remediate this difficulty. This model introduces a
hidden layer of networks based on sparse Gaussian graphical models,
and the observed data are sampled from individual network nodes. In
fMRI, the network layer models the underlying signals of different
brain functional units, and how these units directly interact with
each other. The introduction of this hierarchical structure not only
provides a formal and interpretable approach, but also enables efficient
computation for inferring big networks with hundreds of thousands
of nodes. Based on the conditional convexity of our formulation, we
develop an alternating update algorithm to compute the HGM model parameters
simultaneously. The effectiveness of this approach is demonstrated
on simulated data and a real dataset from a stop/go fMRI experiment.


\noindent \global\long\def\Pr{\mathbb{P}}
 \global\long\def\Ex{\mathbb{E}}
 \global\long\def\mrl{\operatorname{mrl}}
 \global\long\def\ind{\mathbb{1}}
 \global\long\def\var{\operatorname{var}}
 \global\long\def\cov{\operatorname{cov}}
 \global\long\def\argmax{\operatorname{argmax}}
 \global\long\def\diag{\operatorname{diag}}
 \global\long\def\tr{\operatorname{tr}}
 \global\long\def\rank{\operatorname{rank}}

\noindent \global\long\def\boldX{\mathbf{X}}
 \global\long\def\boldZ{\mathbf{Z}}
 \global\long\def\boldSigma{\mathbf{\Sigma}}
 \global\long\def\boldOmega{\mathbf{\Omega}}
\global\long\def\boldG{G}
\global\long\def\boldepsilon{\mathbf{\epsilon}}
 \global\long\def\boldphi{\mathbf{\Phi}}
 \global\long\def\boldmu{\mathbf{\mu}}

\noindent \newpage{}

\section{Introduction}

Graphical models is a statistical tool to describe the relationships
between multiple variables. In functional MRI or fMRI, the variables
are the Blood-Oxygen-Level-Dependent (BOLD) activities at different
regions (also known as voxels) of the human brain. One important scientific
question is how these brain regions are connected, which is termed
brain connectivity. However, the dimensionality of whole-brain fMRI
data is usually in hundreds of thousands, and this scale challenges
direct application of existing methods. Recently, large-scale brain
network estimation is regarded as a ``big data'' problem \citep{turk2013functional},
and has raised several challenges and opportunities \citep{sporns2014contributions}.
In this paper, we propose a novel approach to provide interpretable
and direct estimation from whole-brain fMRI data, and this general
approach may have applications for other big data problems as well.

The fMRI dataset in this paper comes from one subject performing a
cognitive experiment--stop/go trials, which is made publicly available
by its investigators \citep{XueG08}. After preprocessing the data
(see details in Section \ref{sec:fMRI}), it consists of the BOLD
measures from $230,590$ voxels in $180$ time points. Due to the
dimensionality, previous research on stop/go fMRI has been focusing
on individual voxel activations, and their implications for phenotypes
and behavior outcomes. For instance, stop/go voxel activations have
been studied using meta-analysis \citep{SimmondsD08} and causal inference
\citep{Luo12inter}, and the activity predicts substance use and behavior
\citep{ChiangL10,LuoX13}. The problem of studying the whole brain
connections is an emerging and important direction, but is also challenging
methodologically.

There are two major types of connectivity analysis: functional connectivity
and effective connectivity, see \citet{FristonK11}, \citet{SmithS13},
and \citet{BowmanF2013} for review. The simplest measure for inferring
functional connectivity is probably pair-wise correlations. This is
computationally inexpensive but is less biologically meaningful \citep{buckner2010human,SmithS13}.
For example, it does not differentiate direct and indirect connections.
Alternative approaches include clustering \citep{goutte1999clustering,dubois2004identifying},
and independent component analysis or ICA \citep{calhoun2001method,beckmann2004probabilistic,guo2011general,eloyan2013likelihood}.
However, they again lack identification of direct and indirect connections.
Effective connectivity, on the other hand, seeks to infer directional
relationships between directly connected brain regions. The popular
approaches for effective connectivity include dynamic causal modeling
\citep{FristonK03}, structural equation models, and Granger causality.
Because these tools are complex in computation and modeling, they
are usually applied for a small number (e.g. $<100$) of preselected
voxels or regions. The selection criteria include prior knowledge
(e.g. anatomical and functioning annotations), and functional activation
\citep{DuannJ09}. Recently, voxel correlations are used provide more
accurate selection of voxels for a certain region of the brain \citep{zhang2013functional}.
Despite these attempts, the result, however, may be sensitive to the
selection of regions \citep{BowmanF2013}, and the network inference
can be biased if the influence from other omitted regions is large.
Several challenges remain in inferring large-scale direct connectivity
\citep{varoquaux2013learning,sporns2014contributions}. 

To infer large-scale connectivity, we will further develop a popular
type of graphical models, called sparse Gaussian graphical models
\citep{YuanM07,BanerjeeO08}, hereafter sGGM. This type of models
has a solid probabilistic foundation for differentiating direct and
indirect connections \citep{DempsterA72,LauritzeS96}. Suppose we
observe a multivariate observation $\boldX_{n\times p}$ , with $n$
iid observations from a $p$-variate normal distribution $N\left(\boldmu,\boldSigma\right)$.
sGGM represents the relationships between $p$ variables by a network
of $p$ nodes, where each node represents a variable and there are
parsimonious connections between nodes. Methodologically, inference
of the connections between $p$ nodes is reduced to estimate a sparse
inverse covariance $\boldOmega=\boldSigma^{-1}$, where a nonzero
off-diagonal entry in $\boldOmega$ means that the corresponding row
and column variables/nodes are directly connected and a zero entry
means no connections, see \citet{LauritzeS96}. In fMRI applications,
direct connectivity is more biologically interpretable than indirect
functional connectivity \citep{FristonK11}. The sGGM approach also
performs reasonably well on a simulation study using a small number
of regions \citep{SmithS11}.

Recently, several estimators for sGGM have been proposed based on
the $\ell_{1}$ heuristic \citep{Chen1994,TibshiraniR96}. These approaches
can be roughly divided according three major estimation principals:
penalized conditional likelihood or neighborhood selection \citep{MeinshausenN06,YuanM10},
penalized full likelihood or Graphical lasso \citep{YuanM07,BanerjeeO08,FriedmanJ08},
and algebraic properties \citep{CaiT2010,LiuLuo2012}. The last type,
especially, has faster statistical convergence rates for a general
class of distributions with polynomial tails \citep{CaiT2010}, and
is statistically adaptive and computationally efficient \citep{LiuLuo2012}.
However, all these approaches don't scale up to hundreds of thousands
of nodes, though a large scale algorithm for penalized likelihood
is proposed recently by \citet{hsieh2013big}.

\begin{figure}[p]
\begin{centering}
\includegraphics[width=0.9\textwidth]{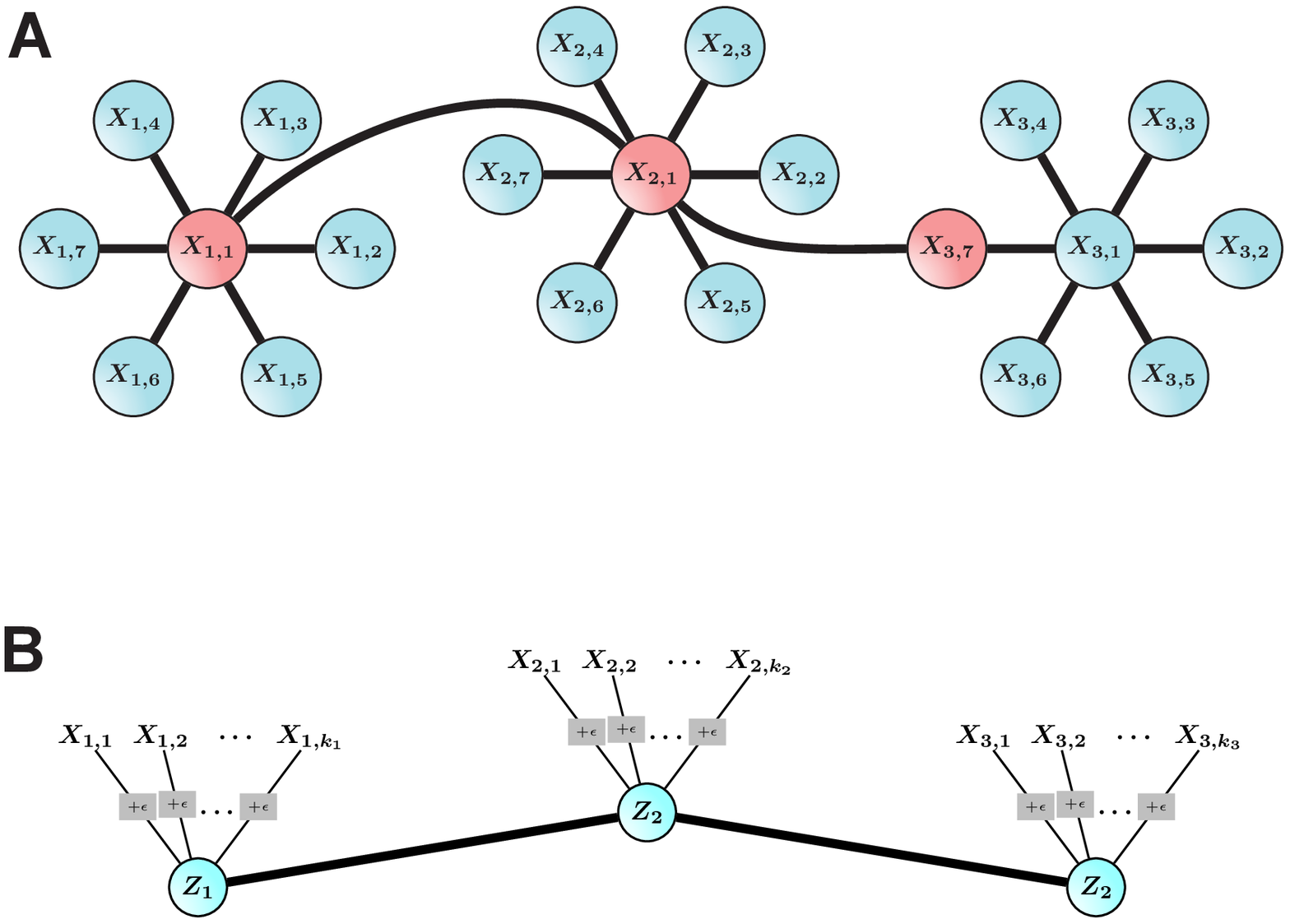}
\par\end{centering}

\caption{\label{fig:modelsketch} Schematic plots of (A) a biological network
model with hubs and communities \citep{guimera2005functional,PowerJ13}
and (B) the HGM model. The biological network model assumes that data
are sampled from all nodes $X_{i,j}$, including the hub nodes $X_{1,0}$,
$X_{2,0}$, and $X_{3,7}$, shown in red. The HGM model assumes a
hidden network of 3 nodes $Z_{i}$, $i=1,2,3$. The observed data
$X_{i,j}$, $j=1,\dotsc,k_{i}$, are the corresponding $Z_{i}$ plus
independent noises. Network nodes are shown in circles, and all the
subscript indexes are arbitrary. }
\end{figure}

The brain network system is rather complex, compared to the standard
sGGM model. For instance, the brain network has a hierarchical structure:
billions of connected neurons excite BOLD measures in hundreds of
thousands of voxels, connected voxels form areas (e.g. motor area),
connected areas form systems (e.g. motor system), and systems interact
with each other. Graph theoretical analysis yields insight understanding
of the network complexity, see \citet{bullmore2009complex} for review.
A popular network structure is shown in Figure \ref{fig:modelsketch}A,
in which communities of nodes are connected with each other via hub
nodes \citep{PowerJ13}. This network structure has also been proposed
for other biological networks, for example genetic networks \citep{guimera2005functional}.
It was unclear how these structures were represented in large-scale
network estimation methods, including sGGM.

Leveraging these scientific findings and methodological advances,
we propose a simple and unified statistical model for big network
data generated in a particular way. A conceptual sketch of this model
is plotted in Figure \ref{fig:modelsketch}B. We employ a hidden layer
of variables to model the network, and the observations are multiple
noisy samples from each node. This model shares similar topological
structures with the hub network (Figure \ref{fig:modelsketch}A),
but incorporate the special characteristics (e.g. smoothness) of fMRI,
see Section \ref{sec:hgm}. Under this framework, we describe three
goals: signal extraction, voxel clustering, and network estimation.
These three goals are interwoven with each other. We thus develop
a generic alternating updating algorithm for carrying out them simultaneous,
see Section \ref{sec:Method}. The advantages of our method  are demonstrated
on the fMRI data in Section \ref{sec:fMRI} and on simulated data
in Section \ref{sec:simu}. We will conclude with discussion in Section
\ref{sec:Discussion}. The technical details are postponed to Section
\ref{sec:Proofs}. 

\section{A Hierarchical Graphical Model\label{sec:hgm}}

We first collect the notations used in this paper. Let $\mathbf{M}=(M_{ij})$
be any matrix. $M_{\cdot j}$ stands for the $j$th column of $\mathbf{M}$,
and $M_{i\cdot}$ for the $i$th row. $ $The following matrix norms
on $\mathbf{M}$ will be used: $\left\Vert \mathbf{M}\right\Vert _{2}=\sqrt{\sum_{i}\sum_{j}M_{ij}^{2}}$
for the Frobenius norm; $\left\Vert \mathbf{M}\right\Vert _{1}=\sum_{i,j}\left|M_{ij}\right|$
for the $\ell_{1}$ norm. The trace and determinant of $\mathbf{M}$
are denoted by $\tr\left(\mathbf{M}\right)$ and $\det\left(\mathbf{M}\right)$
respectively. A diagonal matrix is denoted by $\diag\left(M_{1},\dotsc,M_{p}\right)$
where the diagonal entries $M_{i}$ are shortened notations for $M_{ii}$
for any $i$. The cardinality of a set $S$ is given by $\left|S\right|$.

We now introduce the formulation of our hierarchical graphical model.
Without loss of generality, we assume all variables in our model are
mean centered. This could be achieved by subtracting the mean for
each variable first. In fMRI, the means are usually arbitrary, and
a mean centering approach is usually employed in the preprocessing
pipeline. 

Suppose that \emph{all} the $p$ variables/columns of the observed
data matrix $\boldX$ are separated into one and only one of the $K$
disjoint group sets $G_{k}\subset\left\{ 1,\dotsc,p\right\} $, $k=1,\dotsc,K$.
We introduce a hierarchical variable denoted by $\boldZ$, where each
column $Z_{\cdot k}$ represents the hidden signal for group $G_{k}$.
The hierarchical variable $\boldZ$ relates to the observed variable
$\boldX$ as
\begin{equation}
X_{ij}=Z_{ik}+\epsilon_{ijk}\qquad\mbox{for}\, j\in G_{k},\, i=1,\dotsc,n,\, k=1,\dotsc,K,\label{eq:modelx}
\end{equation}
where the noise variables $\epsilon_{ijk}\sim N\left(0,\phi_{k}\right)$
for group $k$ are independent of each other and independent of all
$\boldZ$ . Denote the error variance matrix $\boldphi=\diag\left(\phi_{1},\dotsc,\phi_{K}\right)$.
As discussed in the introduction, we assume each row of the hierarchical
variable is 
\begin{equation}
Z_{i\cdot}\overset{iid}{\sim}N\left(\mathbf{0},\boldSigma\right)\label{eq:modelz}
\end{equation}
where the inverse covariance (or precision) matrix is $\boldOmega=\boldSigma^{-1}$. 

The observed $\boldX$ under Model \eqref{eq:modelx} and \eqref{eq:modelz}
is still multivariate normal, and thus can be represented by a Gaussian
graphical model. It is possible to directly estimate a $p\times p$
precision matrix $\boldOmega_{X}$ of $\boldX$ for moderate size
$p$, using existing algorithms \citep{FriedmanJ08,CaiT2010,LiuLuo2012}.
This task, however, becomes very challenging in computation and storage
at least, when $p$ is in the size of hundreds of thousands. We will
instead estimate a smaller precision matrix $\boldOmega$ of size
$K\times K$ based on the hierarchical variable $\boldZ$, where $K$
could be much smaller than $p$. The introduction of the hierarchical
variable $\boldZ$ and the group assignment $\boldG$ provides additional
advantages in modeling and interpretation as we will outline below.

The group assignment $\boldG$ makes the resulting network model interpretable
because each node can be interpreted as a functional unit that consists
of several variables. In fMRI, $\boldX$ is the original data where
each column is a voxel, and the voxels in each $G_{k}$ forms a unit
commonly known as area or region of interest (ROI) to neuroscientists.
Our approach allows $\boldG$ to be fixed based on prior knowledge
or estimated from the data by an iterative algorithm in Section \ref{sec:Method}. 

The hierarchical variable $\boldZ$ models the underlying signal within
each group $G_{k}$, and we also use it to model the topological role
of hubs. Given the group assignment $\boldG$, an popular estimate
for $\boldZ$ is given by, for each observation $i$, 
\[
\bar{Z}_{ik}=\frac{1}{\left|G_{k}\right|}\sum_{j\in G_{k}}X_{ij},\quad\mbox{for}\, k=1,\dotsc,K.
\]
This is a common method for extracting signals from ROIs in fMRI analysis.
As we will show momentarily, our estimator of $\boldZ$ is different
from $\bar{\boldZ}$ by incorporating other model parameters, such
as the network $\boldOmega$.

\section{Method\label{sec:Method}}

\subsection{Likelihood Formulation and Convexity}

We introduce the approach to estimate the parameters $\left(\boldZ,\boldG,\boldOmega,\boldphi\right)$
in Model \eqref{eq:modelx} and \eqref{eq:modelz}. We consider two
sGGM algorithms, Glasso \citep{FriedmanJ08} or SCIO \citep{LiuLuo2012}.
Because both are motivated by the penalized likelihood framework,
we will focus on describing the likelihood approach first, while the
difference will emerge later. The negative log-likelihood function
under our hierarchical model is, ignoring constants and scaling factors,
\begin{multline}
L\left(\boldZ,\boldG,\boldOmega,\boldphi\right)=\sum_{k=1}^{K}\left[\sum_{j\in G_{k}}\left\Vert X_{\cdot j}-Z_{\cdot k}\right\Vert _{2}^{2}/\left(n\phi_{k}\right)+\left|G_{k}\right|\log\phi_{k}\right]\\
+\frac{1}{n}\tr\left(\boldZ\boldOmega\boldZ^{T}\right)-\log\det\left(\boldOmega\right)+K\log\left(2\pi\right).\label{eq:like}
\end{multline}
To introduce sparsity on $\Omega$, we consider a LASSO penalty (i.e.
the $\ell_{1}$ norm) formulation via minimizing the following objective
\begin{equation}
L\left(\boldZ,\boldG,\boldOmega,\boldphi\right)+\lambda\left\Vert \boldOmega\right\Vert _{1}\label{eq:penlike}
\end{equation}
where $\lambda>0$ is a penalization parameter. The objective function
\eqref{eq:penlike} is unfortunately not jointly convex in $\left(\boldZ,\boldG,\boldOmega,\boldphi\right)$.
The group assignment $\boldG$ is a combinatorial optimization problem
in general, which is NP-hard. However, the objective is conditionally
convex in all the other parameters, summarized in the following proposition.
\begin{prop}
\label{prop:convex} The objective function \eqref{eq:penlike} is
conditionally convex in $\boldZ$, $\boldOmega$ and $\boldphi^{-1}$
respectively, conditional on all the other parameters in $\left(\boldZ,\boldG,\boldOmega,\boldphi\right)$. 
\end{prop}

\subsection{An Alternating Update Algorithm}

Due to the conditional convexity in Proposition \ref{prop:convex},
we propose to solve the problem via alternating iterative updates
of each parameter, where the updating step minimizes the conditional
objective function. Though the conditional minimization problem is
not convex in the group assignment $\boldG$, similar alternating
procedures have been effective in practice \citep{LloydS82,ForgyE65,MacqueenJ67,HartiganJ79}.
Our algorithm also incorporates a group update step.

The conditional minimization for $\boldZ$ and $\boldphi^{-1}$ are
given in explicit forms in the following proposition. Though the minimization
is over $\boldphi^{-1}$, the solution is conveniently given in terms
of $\boldphi$. The conditional minimization over $\boldOmega$ is
equivalent to the Glasso problem.
\begin{prop}
\label{prop:condmin} The conditional minimizer for $\boldZ$ in \eqref{eq:penlike}
is
\[
\boldZ^{*}=\bar{\boldZ}\mathbf{D}_{G}\left[\mathbf{D}_{G}+\boldOmega\boldphi\right]^{-1}
\]
where $\mathbf{D}_{G}=\diag\left(\left|G_{1}\right|,\dotsc,\left|G_{K}\right|\right)$.
That is, $\boldZ^{*}$ minimizes the following conditional minimization
objective, while $\left(\boldG,\boldOmega,\boldphi\right)$ are fixed,
\[
L_{\boldZ}=\sum_{k=1}^{K}\sum_{j\in G_{k}}\left\Vert X_{\cdot j}-Z_{\cdot k}\right\Vert _{2}^{2}/\left(n\phi_{k}\right)+\frac{1}{n}\tr\left(\boldZ\boldOmega\boldZ^{T}\right).
\]
The conditional minimizer for $\boldphi$ is
\[
\phi_{k}^{*}=\frac{1}{n\left|G_{k}\right|}\sum_{j\in G_{k}}\left\Vert Z_{\cdot k}-X_{\cdot j}\right\Vert _{2}^{2},\quad\mbox{for}\, k=1,\dotsc,K,
\]
where the conditional objective is
\[
L_{\boldphi^{-1}}=\frac{1}{n}\sum_{k=1}^{K}\left[\phi_{k}^{-1}\sum_{j\in G_{k}}\left\Vert X_{\cdot j}-Z_{\cdot k}\right\Vert _{2}^{2}-\left|G_{k}\right|\log\phi_{k}^{-1}\right].
\]
The conditional minimization problem of $\boldOmega$ is equivalent
to the Glasso objective
\[
L_{\boldOmega}=\frac{1}{n}\tr\left(\boldZ\boldOmega\boldZ^{T}\right)-\log\det\left(\boldOmega\right)+\lambda\left\Vert \boldOmega\right\Vert _{1}.
\]

\end{prop}
We use the conditional minimizers for $\boldZ$ and $\boldphi$ as
our iterative updates respectively. Because the purpose of minimizing
$L_{\boldOmega}$ is to produce a sparse precision matrix $\boldOmega$,
we consider two approaches, Glasso and SCIO. Glasso minimizes $L_{\boldOmega}$
exactly, and SCIO is based on the algebraic properties derived from
$L_{\boldOmega}$ \citep{CaiT2010,LiuLuo2012}, which has faster convergence
rates for $\boldOmega$ under moderate distribution assumptions (e.g.
heavier tails). Because SCIO does not enforce positive definiteness
of the precision matrix, we perform a simple refitting approach to
ensure such \citep{CaiT2010}. 

The update rule for $\boldZ$ is a linear combination of $\bar{\boldZ}$,
where the combination depends on other parameters. Due to the shrinkage
effect by the hierarchical variable \citep{LehmannE98}, our estimate
is expected to have smaller MSEs than $\bar{\boldZ}$.

The algorithm for solving our hierarchical graphical model problem
is summarized in Algorithm \ref{alg:generic}. The convergence criterion
for stopping the iterative updates is
\begin{equation}
\frac{\left\Vert \boldZ^{\left(t-1\right)}-\boldZ^{\left(t\right)}\right\Vert _{2}}{\max\left(1,\left\Vert \boldZ^{\left(t-1\right)}\right\Vert _{2}\right)}<e_{\mbox{tol}}\quad\mbox{and}\quad\boldG^{\left(t\right)}=\boldG^{\left(t-1\right)}\label{eq:converge}
\end{equation}
where $e_{\mbox{tol}}$ is a tolerance level (e.g. $10^{-4}$), $\boldZ^{\left(t\right)}$
is the update at iteration $t$, and similar definition for $\boldG^{\left(t\right)}$,
$\boldOmega^{\left(t\right)}$, and $\boldphi^{\left(t\right)}$. 

There are many ways to update $\boldG$ in step 4 of Algorithm \ref{alg:generic}.
For simplicity, we use a hybrid rule for finding the group assignment
$\boldG$. In the initialization stage, we set $\left(\boldZ^{\left(0\right)},\boldG^{\left(0\right)}\right)$
from Hartigan's k-means \citep{HartiganJ79} because it usually provides
good clustering in practice. We then use $\left(\boldZ^{\left(0\right)},\boldG^{\left(0\right)}\right)$
in step 2 and 3 to initialize $\boldOmega^{\left(0\right)}$ and $\boldphi^{\left(0\right)}$.
For the sequential alternating updates, we use a simple assignment
rule suggested by \citet{LloydS82}, \citet{ForgyE65}, and \citet{MacqueenJ67},
where each point is reassigned to the closest cluster center $Z_{\cdot k}$.
Because k-means suffers from the issue of converging to a local optima,
our alternating algorithm may also suffer from this issue. Thus, we
consider multiple runs of our algorithm and select the one with the
largest likelihood.

\begin{algorithm}[p]
\textbf{Initialize}: $ $$\left(\boldZ^{\left(0\right)},\boldG^{\left(0\right)},\boldOmega^{\left(0\right)},\boldphi^{\left(0\right)}\right)$,
$t=0$.

\textbf{Repeat until the convergence criterion \eqref{eq:converge}
is met:}
\begin{enumerate}
\item Given $\left(\boldZ^{\left(t\right)},\boldG^{\left(t\right)},\boldOmega^{\left(t\right)},\boldphi^{\left(t\right)}\right)$,
update $\boldZ^{\left(t+1\right)}=\bar{\boldZ}\mathbf{D}_{G^{\left(t\right)}}\left[\mathbf{D}_{G^{\left(t\right)}}+\boldOmega^{\left(t\right)}\boldphi^{\left(t\right)}\right]^{-1}$. 
\item Given $\left(\boldZ^{\left(t+1\right)},\boldG^{\left(t\right)},\boldOmega^{\left(t\right)},\boldphi^{\left(t\right)}\right)$,
update $\phi_{k}=\frac{1}{n\left|G_{k}\right|}\sum_{j\in G_{k}}\left\Vert Z_{\cdot k}-X_{\cdot j}\right\Vert _{2}^{2}$
for $k=1,\dotsc,K$. 
\item Given $\left(\boldZ^{\left(t+1\right)},\boldG^{\left(t\right)},\boldOmega^{\left(t\right)},\boldphi^{\left(t+1\right)}\right)$,
update $\boldOmega^{\left(t+1\right)}$ by a precision matrix estimation
method (either Glasso or SCIO).
\item Given $\left(\boldZ^{\left(t+1\right)},\boldG^{\left(t\right)},\boldOmega^{\left(t+1\right)},\boldphi^{\left(t+1\right)}\right)$,
update $\boldG^{\left(t+1\right)}$ such that each $X_{\cdot j}$
is re-assigned to the closest center $Z_{\cdot k}$.
\item Update $t=t+1$.
\end{enumerate}
\caption{\label{alg:generic} An alternating update algorithm for estimating
the HGM parameters. }
\end{algorithm}

\subsection{Choice of the Tuning Parameters }

Our model contains two tuning parameters, $K$ and $\lambda$. These
can be chosen using either existing scientific knowledge or model
selection methods. We employ the scientific choice in Section \ref{sec:fMRI},
and here we describe a model selection approach when such scientific
knowledge is not available. The Bayesian Information Criterion (BIC)
for our model is
\[
L_{K,\lambda}+\frac{\log p}{n}\left(s/2+p+K\left(n+2\right)-1\right)
\]
where $L_{K,\lambda}$ is the negative log-likelihood \eqref{eq:like}
with the choice of $K$ and $\lambda$, evaluated at the converged
solution produced by Algorithm \ref{alg:generic}, and $s$ is the
number of nonzeros in the off-diagonals of the solution $\boldOmega$.
The model complexity component in above consists of those of k-means
\citep{pelleg2000x} and sGGM. That is, there are $K-1$ class probabilities
from $\boldG$, $K$ variance estimates from $\boldphi$, $Kn$ estimates
from $\boldZ$, $s/2+p$ nonzeros from $\boldOmega$. The tuning parameter
$\lambda$ controls the number of nonzeros in $\boldOmega$, and one
can perform a grid search on $\lambda$ first to pick the value yielding
the smallest BIC. The tuning parameter $K$ controls the number of
groups. One then compares the minimal BIC values from the previous
step with different choices of $K$, and chooses the $K$ that produces
the smallest BIC.

\section{A fMRI Study on Motor Prohibition\label{sec:fMRI}}

We use an fMRI dataset \citep{XueG08} to illustrate the effectiveness
of HGM. The dataset is publicly available from Open fMRI (https://openfmri.org/data-sets)
under the accession number ds000007. This whole dataset consists of
20 subjects scanned under several sessions, with different kinds of
stop/go event tasks. For the illustration purpose, we analyze the
session 1 data of subject 1.

As suggested by the authors \citep{XueG08}, we employ the same preprocessing
pipeline implemented in the FMRIB software library (FSL, available
from http://fsl.fmrib.ox.ac.uk/fsl/fslwiki/). Briefly, the pipeline
includes slice timing correction, alignment, registration, normalization
to the average 152 T1 MNI template, and smoothed with a 5mm full-width-half-maximum
Gaussian kernel. The data are denoised using the FSL MELODIC procedure
and a high pass filter with a 66s cut-off. After preprocessing, general
linear models (GLM) for each voxel \citep{Friston1K99} are used to
remove the non-stationary components, including motion and event related
activation. The standardized GLM residuals are retained for our HGM
analysis. The residuals are assumed to be stationary, similar to resting-state
fMRI \citep{FairD07}.

The processed dataset consists of the residual BOLD activity from
230,590 voxels in 180 time points, or equivalently an input matrix
$X$ with $n=180$ and $p=230,590$. To make each variable in $X$
comparable, we standardize each one to have mean zero and unit variance.
To illustrate the complex networks that can be recovered by HGM, we
fix $K=200$ and $\lambda=0.5$, because these choices roughly matches
the usual number of brain parcellations, for example the AAL atlas
\citep{tzourio2002automated}. The resulting network has interesting
scientific interpretations as well. Other choices can be taken depending
on the scientific goals. For example, larger $K$ will give finer
parcellations of the brain, smaller $\lambda$ usually yields densely
connected networks, and vice versa. 

\begin{figure}[p]
\begin{centering}
\includegraphics[width=0.9\textwidth]{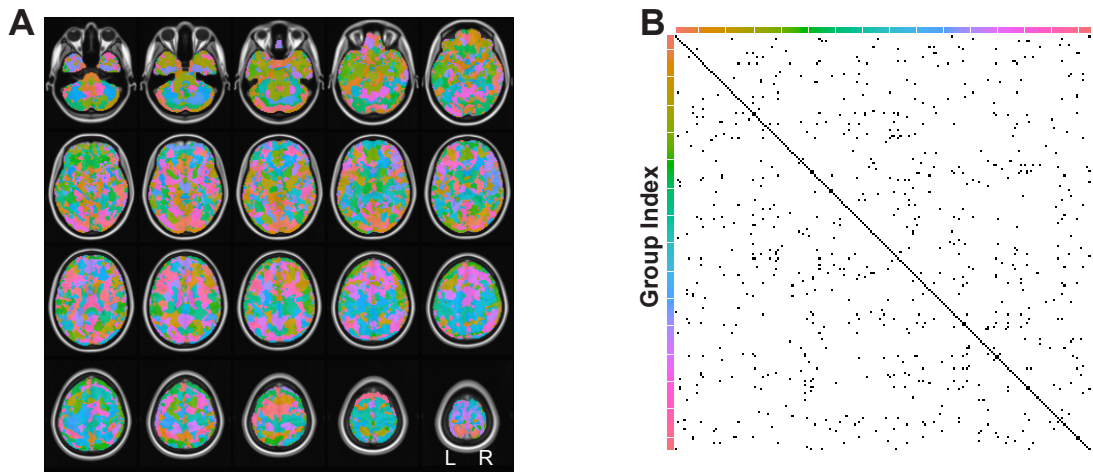}
\par\end{centering}

\caption{\label{fig:hgm-whole}Recovered voxel groups (A) and network edges
(B) by HGM. (A) voxel groups are indexed by color, overlaid on the
average 152 T1 MNI template. (B) Connections between two groups are
shown in black entries in the matrix, where the two groups are given
by the row and column index colors respectively; no connections are
shown in white. L: left; R: right.}

\end{figure}

We will use the SCIO update in Algorithm \ref{alg:generic}, because
it shows higher accuracy for recovering the network edges in our simulation
study (Section \ref{sec:simu}). To avoid local minima, due to group
assignments, our HGM algorithm (Algorithm \ref{alg:generic}) is repeated
10 times with random starts. More number of repeats are allowed if
more computing resources are available, but we find that 10 repeats
are sufficient and the parcellations are stable, see Section \ref{sec:simu}
for a simulation evaluation as well. The repeat with the largest likelihood
is reported in Figure \ref{fig:hgm-whole}. To a certain extent, the
voxel grouping shows symmetry between left and right hemispheres,
though this is not imposed in HGM. This approximate bilateral symmetry
coincides with the classic theory of (approximately) mirrored functions
of the two hemispheres. We thus are inclined to postulate that the
HGM grouping recovers different functional units. It is challenging
to visualize all the resulting voxel groups and their connections
in Figure \eqref{fig:hgm-whole}B, and we thus examine one group and
its connections in detail.

The region preSMA (anterior part of supplementary motor area) has
been shown to play an important role during stop/go trails \citep{DuannJ09},
and it is tagged as group 1 by HGM. Note that the group index numbers
are arbitrary. We examine the overlays of big voxel clusters ($\ge100$
voxels) in group 1 and three other directly or indirectly connected
groups, see Figure \ref{fig:hgm-net}. The detailed coordinates and
cluster sizes are included in the supplemental material of this paper.
Each group may contain multiple regions clustered together, and most
of them are closely located in the brain and mirrored on both left
and right hemispheres. The specific clusters in HGM also provides
new insight about the wiring around preSMA, partly because HGM performs
simultaneously brain parcellation and direct connectivity estimation.
For instance, the connection between preSMA and rIFC (right inferior
frontal cortex) have been studied before using a small Granger model
\citep{DuannJ09}, and here the whole brain HGM suggests that this
connection is direct even if considering the whole brain activity.
Moreover, insula activation has been implicated in a previous stop/go
study \citep{LuoX13}, and it also correlates with preSMA \citep{zhang2012resting}.
HGM provides a more detailed map of anterior insula and rIFC (in group
200), which is directly connected to the preSMA group (group 1). This
connection is termed the salience network \citep{seeley2007dissociable}.
The inclusion of rIFC and the exclusion of other salience network
regions prompt an interesting question how the connections within
this network are wired, for example whether the insula correlation
is a consequence or cause of the direct connectivity between rIFC
and preSMA. Furthermore, our HGM result also suggests that the insula
grouping depends on the anterior (group 200) and posterior (group
65) positions, consistent with the findings from several studies \citep{anderson2011connectivity,jakab2012connectivity,kelly2012convergent}.
Finally, the preSMA group contains two distant regions, preSMA and
superior part of cuneus (or Brodmann area 19, superior), which is
probably due to the role of the latter in motion-related visual processes.
These two regions are also clustered by an ICA study \citep{sauvage2011reevaluating}.
These coherent results suggest that further investigation is needed
to study the possible connection between these two regions.

\begin{figure}[p]
\begin{centering}
\includegraphics[width=0.9\textwidth]{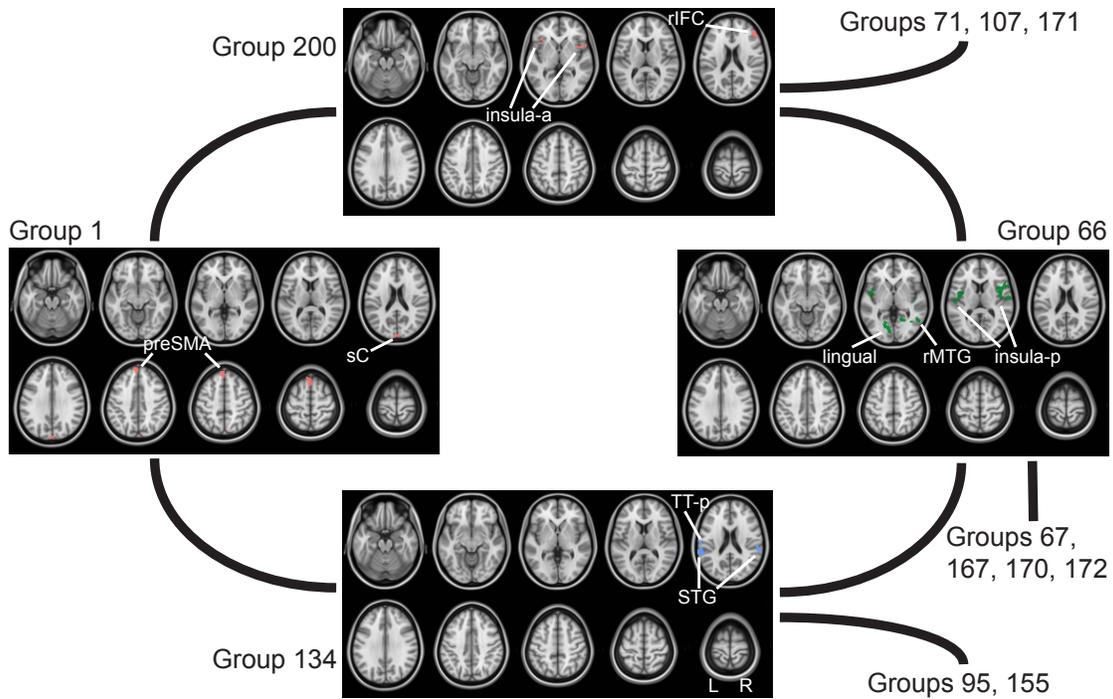}
\par\end{centering}

\caption{\label{fig:hgm-net} The overlays of voxel groups on the average 152
T1 MNI template, for the preSMA group (group 1) recovered by HGM,
two directly connected groups (134, 200), and one group (group 66)
connected to the previous two groups. The group index numbers are
arbitrary. preSMA: anterior part of supplementary motor area; rMTG:
right middle temporal gyrus, or Brodmann area 22; insula-a: anterior
insula; insula-p: posterior insula; rIFC: right inferior frontal cortex;
sC: superior cuneus, or Brodmann area 19; STG: superior temporal gyrus;
TT-p: posterior transverse temporal, or Brodmann area 42; L: left;
R: right.}
\end{figure}

\section{Simulations\label{sec:simu}}

We assess the performance of HGM using the following simulation model.
The hierarchical variable $\boldZ$ are $180$ iid samples from mean
zero multivariate normal with a $200\times200$ precision matrix $\boldOmega^{*}$.
The precision matrix $\boldOmega^{*}$ is block diagonal with block
size $5$ where each block has off-diagonal entries equal to $0.8$
and diagonal $1$. The order of nodes are then randomly permuted,
and the covariance matrix is scaled such that all the marginal variances
equal to 1. A similar model has been used before \citep{LiuLuo2012}. 

In each simulation run, each column of $\boldZ$ is added by 50 iid
standard normal vectors respectively. This yields the observed matrix
$\boldX$ of dimension $180\times10,000$, with signal to noise ratio
1. The simulation parameters ($n=180$, $p=10,000$, $K=200$) are
similar to the scale of the fMRI data. All simulations are repeated
50 times.

\subsection{Network Estimation}

\begin{figure}[p]
\begin{centering}
\includegraphics[width=0.9\textwidth]{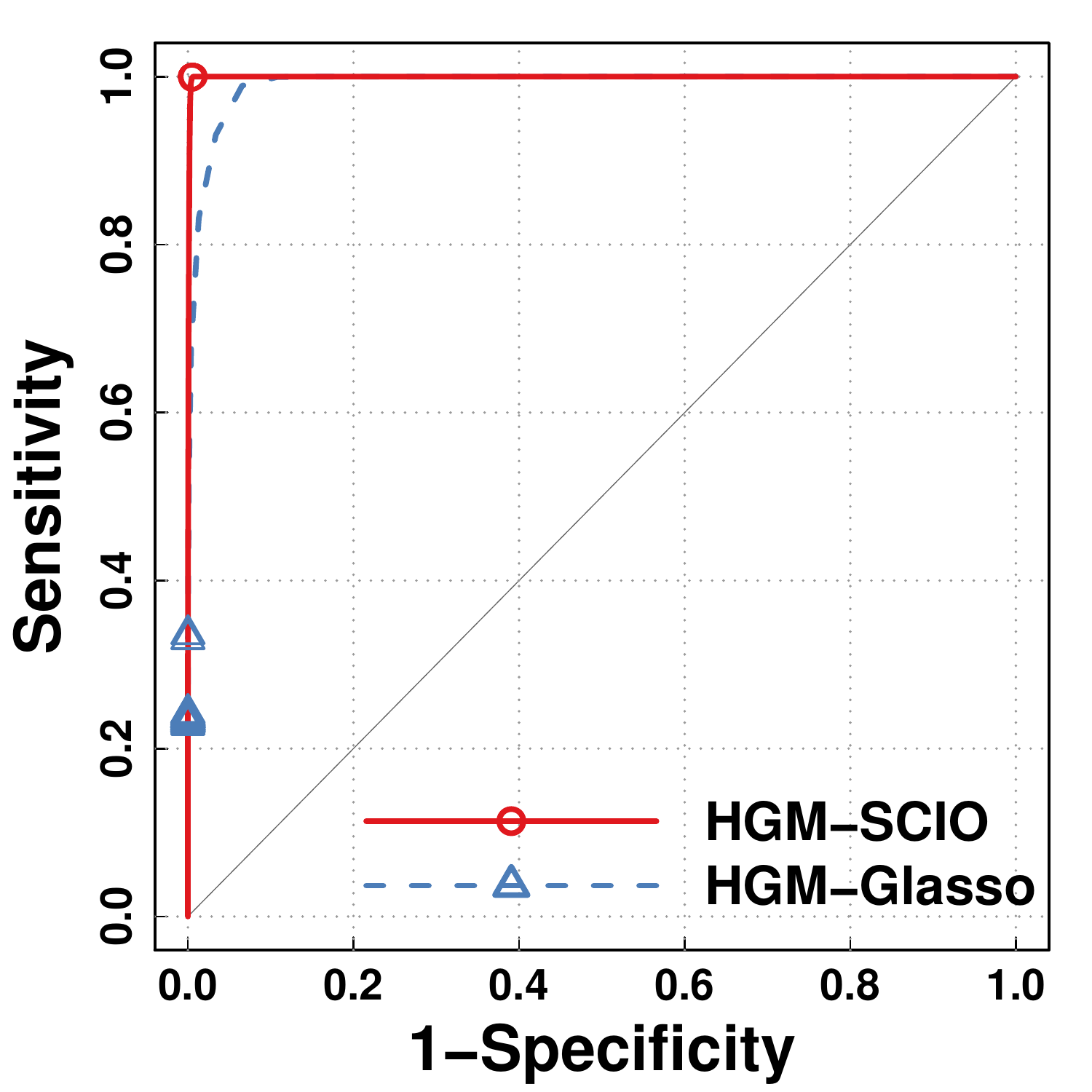}
\par\end{centering}

\caption{\label{fig:simuNet} Average receiver operating characteristics on
estimating the network edges by embedding SCIO (red solid line) and
Glasso (blue dash line) estimators in the HGM algorithm, averaged
across 50 runs. The sensitivity and specificity values with the BIC
selected $\lambda$ are overlaid by red circles (HGM-SCIO) and blue
triangles (HGM-Glasso). HGM-SCIO: HGM with the SCIO estimator in Algorithm
\eqref{alg:generic}; HGM-Glasso: HGM with Glasso.}
\end{figure}

It is difficult compare results with different group assignments $\boldG$.
We first consider the true $\boldG$ is given and fixed in our HGM
algorithm. That is, we initialize with $\boldG^{\left(0\right)}=\boldG$,
and we don't perform the group update step in Algorithm \ref{alg:generic}.
We compare two methods to estimate the precision matrix $\boldOmega$,
SCIO and Glasso. Both methods contain a tuning parameter $\lambda$
controlling the sparsity of the matrix, and grids of $50$ values
are used in both methods. The network edge identification accuracy
with varying $\lambda$ is assessed by the average receiver operating
characteristics (ROC), see Figure \ref{fig:simuNet}. Overall, the
HGM models with both methods, named as HGM-SCIO and HGM-Glasso, have
good performance, and HGM-SCIO clearly outperforms HGM-Glasso. When
the tuning parameter $\lambda$ is chosen by BIC for both methods,
HGM-SCIO has higher sensitivity than HGM-Glasso, while maintaining
high specificity as well.

\subsection{Group Estimation}

\begin{figure}[p]
\begin{centering}
\includegraphics[angle=90,width=0.9\textwidth]{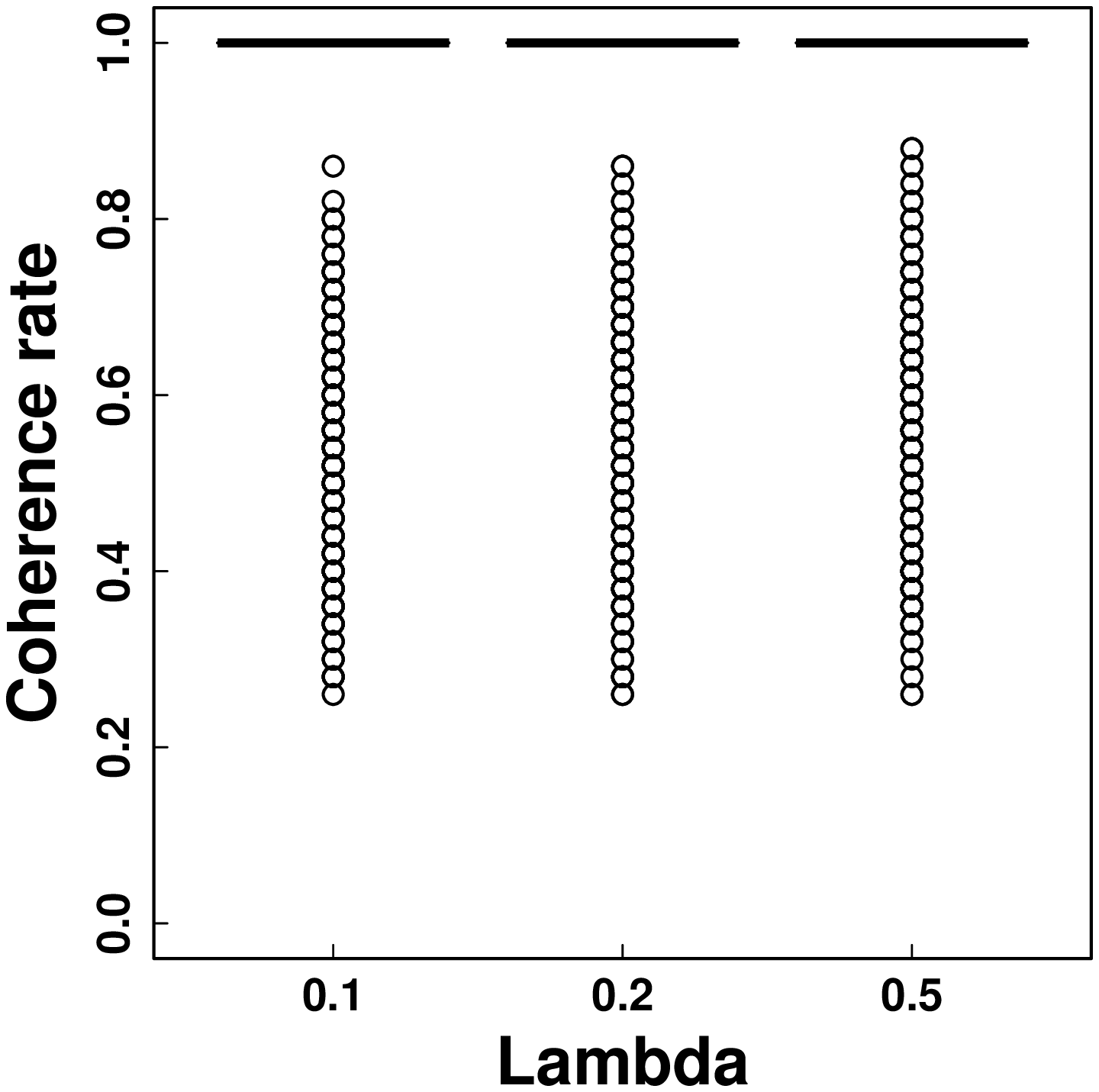}
\par\end{centering}

\caption{\label{fig:simuGroup} Boxplots of the coherence rates with varying
$\lambda$ over 50 runs for all groups. Over $89\%$ of the coherence
rates equal to 1 exactly across all choices of $\lambda$. }
\end{figure}

To assess the group assignment accuracy, we run all the steps in Algorithm
\ref{alg:generic} on the simulated data. To fix ideas, we consider
3 values for $\lambda$: $0.1$, $0.2$, and $0.5$. The resulting
network connections are dense, moderate, and sparse respectively.
For comparing each estimated $\hat{\boldG}$ with the true group $\boldG$,
we use the following simple measure, coherence rate, 
\[
r_{k}=\frac{\max_{i}\left|\hat{G}_{i}\cap G_{k}\right|}{\left|G_{k}\right|},\quad k=1,\dotsc,K.
\]
For each run, we initialize with 10 different initial group assignments,
and retain the estimated $\hat{\boldG}$ with the largest likelihood.
Due to the symmetry of our simulated groups, we pool the coherence
measures from different groups across 50 runs. Figure \ref{fig:simuGroup}
shows that, by the coherence measure, HGM is stable and accurate in
group estimation with varying $\lambda$, and $89\%$ of the group
estimates equal to the truth exactly.%

\section{Discussion\label{sec:Discussion}}

This paper introduces a model for estimating direct connections from
large-scale data, motivated by the whole-brain network estimation
problem in fMRI. We aim to provide a computationally efficient and
interpretable model for data with hundreds of thousands of variables.
We propose an alternating update algorithm to estimate the model parameters
simultaneously. The model is illustrated using both simulated data
and a dataset from a stop/go fMRI experiment. 

The interpretation of connections recovered by HGM, however, should
be treated with caution. It has been well know than fMRI BOLD signals
are confounded by varying hemodynamic processes across the brain,
and thus the connectivity should interpreted on the bold level. Moreover,
the spatial resolution of fMRI is insufficient for inferring neuronal
connections. Thus the connections in HGM should not simply treated
as direct neuronal connections, though these can be highly correlated.

There are several possible hypotheses for the voxel grouping in HGM.
One may conjecture that smoothness may play a part, and thus nearby
voxels are grouped together. However, this simplified explanation
does not explain the long-range connections in HGM. We are inclined
to hypothesize that the grouping is mostly due to close brain functions.
Certainly nearby voxels may share similar brain functions. By grouping,
HGM provides two levels of interpretation: how groups are connected
to each other, and how brain regions are grouped. It will be interesting
to investigate the biological and graph theoretical foundations of
these two levels. 

The theoretical aspects of HGM are not developed here. Though finite
sample theory exists for the precision matrix estimation \citep{CaiT2010,LiuLuo2012},
the assumptions are difficult to verify in big data collected from
biological experiments (e.g. fMRI). Furthermore, the group assignment
is a combinatorial problem, which is difficult to analyze. We, instead,
provide the HGM model as a first step to understand large brain networks. 

The distribution assumptions can be relaxed. Even if the distribution
of $\boldZ$ has heavier tails, the same convergence rates hold \citep{CaiT2010,LiuLuo2012}.
More general distributions can be accommodated by nonparametric covariance
estimates \citep{LaffertyJ2012sparse}, which will be interesting
to explore in the future. 

The independent assumption in $\boldZ$ can also be replaced with
a matrix normal distribution \citep{LengC12} to describe temporal
dependence. In our application example, the whitening step in preprocessing
may reduce such dependence. It is also interesting to develop spatio-temporal
models to validate this assumption, such as \citet{kang2012spatio}. 

Several extensions of HGM are possible. For instance, one may consider
modeling the probability of group assignments for each voxel. One
may consider studying the group-level HGM from multiple subjects and
sessions, either by embedding HGM in mixed models or by the group
Lasso penalty \citep{YuanM06b}. One may also consider incorporating
the distance between voxels to help group assignments, but it should
be pointed out that the choice of metric may be challenging. For example,
Euclidean distance is usually not a good choice of metric for voxels
\citep{BowmanF2013}. We will leave these directions to future research.

\section{Proof\label{sec:Proofs}}

The proof of Proposition \ref{prop:condmin} implies the proof of
Proposition \ref{prop:convex}, and thus the latter is omitted.

The terms in $L$ that are relevant to $\boldZ$ are, ignoring other
factors irrelevant of $\boldZ$, 
\[
L_{\boldZ}=\sum_{k=1}^{K}\sum_{j\in G_{k}}\left\Vert X_{\cdot j}-Z_{\cdot k}\right\Vert _{2}^{2}/\left(n\phi_{k}\right)+\frac{1}{n}\tr\left(\boldZ\boldOmega\boldZ^{T}\right).
\]
The derivative of $L_{\boldZ}$ with respect to $\boldZ$ is proportional
to
\[
\boldZ\mathbf{D}_{G}\boldphi^{-1}+\boldZ\boldOmega-\bar{\boldZ}\mathbf{D}_{G}\boldphi^{-1}.
\]
Multiplying the derivative by $\boldphi$, solve $\boldZ$ that sets
the product zero to yield the minimizer
\[
\boldZ=\bar{\boldZ}\mathbf{D}_{G}\left[\mathbf{D}_{G}+\boldOmega\boldphi\right]^{-1}.
\]

Similar to the derivation of the minimizer for $\boldZ$, the conditional
objective function is
\[
L_{\boldphi^{-1}}=\frac{1}{n}\sum_{k=1}^{K}\left[\phi_{k}^{-1}\sum_{j\in G_{k}}\left\Vert X_{\cdot j}-Z_{\cdot k}\right\Vert _{2}^{2}-\left|G_{k}\right|\log\phi_{k}^{-1}\right].
\]
the derivative with respect to $\phi_{k}^{-1}$ equals to, for every
$k$,
\[
\frac{1}{n}\sum_{j\in G_{k}}\left\Vert X_{\cdot j}-Z_{\cdot k}\right\Vert _{2}^{2}-\left|G_{k}\right|\phi_{k}.
\]
The solution that sets the derivative to zero is, for every $k$,
\[
\phi_{k}=\frac{1}{n\left|G_{k}\right|}\sum_{j\in G_{k}}\left\Vert Z_{\cdot k}-X_{\cdot j}\right\Vert _{2}^{2}.
\]

Finally, the terms relevant to $\boldOmega$ in $L$ are
\[
L_{\boldOmega}=\frac{1}{n}\tr\left(\boldZ\boldOmega\boldZ^{T}\right)-\log\det\left(\boldOmega\right)+\lambda\left\Vert \boldOmega\right\Vert _{1},
\]
and the minimization is thus equivalent to a Glasso problem on $\boldOmega$.

\bibliographystyle{apa}
\bibliography{../../rossi}

\end{document}